# Citadel E-Learning: A New Dimension to Learning System

[1]Awodele O., [2]Kuyoro S. O., [3]Adejumobi A. K., [4]Awe O. and [5]Makanju O.
Department of Computer Science and Mathematics
Babcock University
Ilishan-Remo, Ogun State, Nigeria

Abstract- E-learning has been an important policy for education planners for many years in developed countries. This policy has been adopted by education in some developing countries; it is therefore expedient to study its emergence in the Nigerian education system. The birth of contemporary technology shows that there is higher requirement for education even in the work force. This has been an eye opener to importance of Education which conveniently can be achieved through E-learning. This work presents CITADEL E-learning approach to Nigeria institutions; its ubiquity, its implementations, its flexibility, portability, ease of use and feature that are synonymous to the standard of education in Nigeria and how it can be enhanced to improve learning for both educators and learners to help them in their learning endeavour.

## I. INTRODUCTION

E-learning is an innovative approach for delivering electronically mediated, well-designed, learner-centred interactive learning environments to anyone, anyplace, anytime by utilizing the internet and digital technologies in respect to instructional design principles. It involves learning through the use of ICT infrastructures. In this age, learning through the use of computer is simply online way of acquiring knowledge through the internet. The online learning involves the use of Internet Browser or Navigator. It may be in form of Audio, Visual, and or Audio-Visual. The convergence of the internet and learning, or Internet-enabled learning is called E-learning.[5]

In Nigeria, very few universities are currently carrying out their academic activities through one form of ICT or the other, while the urge to embark on E-learning is still a far-fetched dream to some, because their ICT infrastructure is very inadequate. The rapid expansion of ICTs in Nigeria offers an opportunity to consider its use in the promotion of distance education. It offer students considerable benefits including increase access to learning opportunities, convenience of time, and place, making available a greater variety of learning resources, improve opportunities for individualized learning and emergence of more powerful cognitive tools.

In most developing countries, the percentage of resources available for students is very minimal, such that the tradition mode of learning, limits the student's assimilation and their knowledge only to available resources. With E-learning; resources such as course materials, and other learning enhanced instructional will be unlimited and would be at the fingertips of the students. Very few of the E-learning environments lay emphasis on the importance of motivation for students and lecturers alike, in sense that their non user-friendly and hectic environments reduces motivation of both the students and lecturers. This has been considered and thus, incorporated in the project by making it interactive and user-friendly. Learning environment that support administration functionality, feedback functionality will be adopted. The CITADEL e-learning environment allows easy access that will motivates Instructors and students to participate effectively.

## II. RELATED WORK

There are various literatures on E-learning, some of which has been carefully reviewed in this work. Learning environments such as Sakai Project and Blackboard Learning System are considered. Traditional academic learning as teacher-centred instruction of synchronous and scheduled groups, constrained by classroom availability, while e-learning is student-centred, asynchronous, and available anytime and anywhere. E-learning can be both highly interactive and simultaneously isolating because of the inherent difficulties of developing cohesiveness and true connectedness among students (Sauer 2001). Unfortunately, few research focus on the use and effectiveness of e-learning processes in Nigeria.

[6] emphasized the attention given to e-learning and its development. There has been business and academic interest in e-learning; views such as the perspective of brokers, educational service and content providers were considered. E-learning was viewed according to its basic aspect namely:

Content, Services, Agents and Rules. The use of well structured descriptions of content material has contributed to the desired materials. Mentioned also is the agent playing role in E-learning environment. These agents are Active learner, collaborative learners, passive learners, content developers and administrators. It was therefore concluded that agents and roles should also be incorporated in the web service definitions and eventually can be accessed by the individuals by related metadata that have been registered with an online web service registry and then the registry can be searched manually or programmatically to discover and select components.

[3] explicitly elaborates the importance and challenges faced by the user of blackboard. The benefits include quick feedback, increased availability, improved communication, tracking and skill building. It was also stated that blackboard can be use as supplement to classroom learning even when other digital environment learning system are the primary instructional tool. He went further to enumerate the drawbacks which are hardness to learn and non-flexibility. A survey of staff and student in University of Wisconsin system who majorly use blackboard management system find it difficult to learn because of its non-flexibility. The study also found that despite expectations, many students were not proficient with the technology. A separate study, an evaluation of Blackboard as a platform for distance education delivery at Hampton University School of Nursing, found that the internet is often a new learning environment for those returning to University for graduate degrees. Result showed that blackboard users are provided with course management system that delivers learning content and resources to them.

[4] shows in his study that Sakai Project, an open source collaborative learning environment, has been adopted in different institution around the globe. University of Fernando Pessoa (UFP) started a pilot experience with Sakai 1.0 in October 2004 opening it to all instructors and students alike . One year later, around 782 users had logged in and 150 sites were active since then the growth of Sakai has increased.

Having carefully analysed some literature on e-learning environments, it is discovered that blackboard management system is not user-friendly and it is non-flexible and other e-learning management systems have been adopted in various countries because they are applicable to the environment. A new System called CITADEL e-learning environment that conforms to the education system of Nigeria is developed in this work. The usual student identification/matriculation number can be used on this system. It provides features as those in the traditional learning but ubiquitously. As such, it can be implemented effectively in any tertiary institution in Nigeria. It is not platform dependent therefore it can be implemented on any operating system.

## III. PROPOSED SYSTEM

This work is designed to accentuate the flexibility and simplicity of e-learning in Nigerian institutions. CITADEL e-learning system is a software package for producing Internet-based courses. It is a global development project designed to support a social constructionist framework of education. Based on evaluation of all available web development tools, HTML, ADOBE FIREWORKS was used for the interface of the environment, PHP for interaction with the database, JAVASCRIPT for more advanced interactive interface and MYSQL as the database to keep track of data are used to develop the software to make CITADEL effective and easily adaptable in Nigeria. The system is designed with flexibility in mind; and consideration for expansion was paramount in the design. Other factors considered in the design include security, portability, usability and reusability. In general, it was ensured that the design conforms to the ISO standard.

The system architecture is in three different layers. These layers are responsible for the presentation of information from the back-end of the system to the users, monitors the interaction between the data and user and the also provides control on the DBMS. The layers are as follows

- Presentation layer
- Middle-tier layer
- Data layer

The listed layers are represented below in Fig. 1

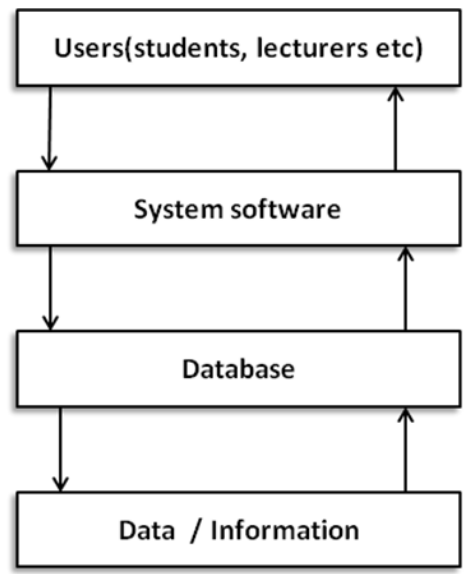

Figure 1: System Architecture

## IV. IMPLEMENTATION

Citadel E-learning environment comprises all forms of electronically       supported learning and teaching, which are procedural in character and aim to effect the construction of knowledge with reference to individual experience, practice and knowledge of the learner. Information and communication systems, serve as specific media to implement the learning process. CITADEL is essentially the use of computer and network-enabled transfer of skills and knowledge. Content is delivered via the Internet, audio or video tape, and CD-ROM. It includes media in the form of text, image, animation, streaming video and audio. It has the functionality such as to create and deliver courses through media such as videos and audio as well as document courses such as PDFs, grades and access learning outcomes (continuous assessment), manage students, courses and resources and also to track student and course progress from standard administrative report. It has three modules. These modules include:

- **Student:** This is the student page where he or she can login with password and username. Once access is granted the student can access the various features available on the page. This includes: courses, registered courses, downloads, lectures, assignment, examinations and quizzes, results, timetable etc. The index page is the page through which the student would gain access into the system. On the page is a login form with two input boxes provided for username (Matric number) and password. When this data is supplied the system validates the data, if the data supplied is valid the student gains entry into the system else access is denied as shown in Fig. 2.

- **Lecturer:** This where the lecturers also login in with password and access functionalities such as post assignment, upload lectures, create examinations, start alive class through video conferencing etc. The lecturer login page as depicted in Fig. 3 is the page from which the lecturer would use in order to gain access into the lecturer window which will take him to his home page. In this window the lecturer can choose various tasks like preparing quiz, lecture notes, give assignment etc.

- **Registry:** The registry performs the same function as those of traditional learning environment. The function of the registry is to register students, lecturer, create department and faculties. This section uses screenshots to show the features and use of CITADEL learning environment and elaborate on the three modules of the software. It also shows it documentation and step by step user guide. The registrar login page shown in Fig. 4 is the page from which the registry would use in order to gain access into the registry window and perform all the functionality available to page.

Other windows such as Messages, View Classmates, Assignment, Exam and quizzes, timetable, Library, Downloads, View syllabus, View Lecturers, Notice, Chat, etc are depicted at the appendix.

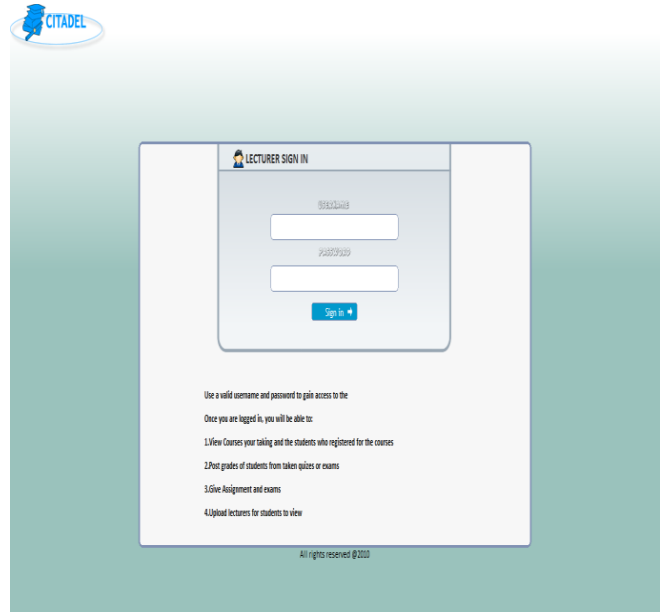

Figure 2: Index page

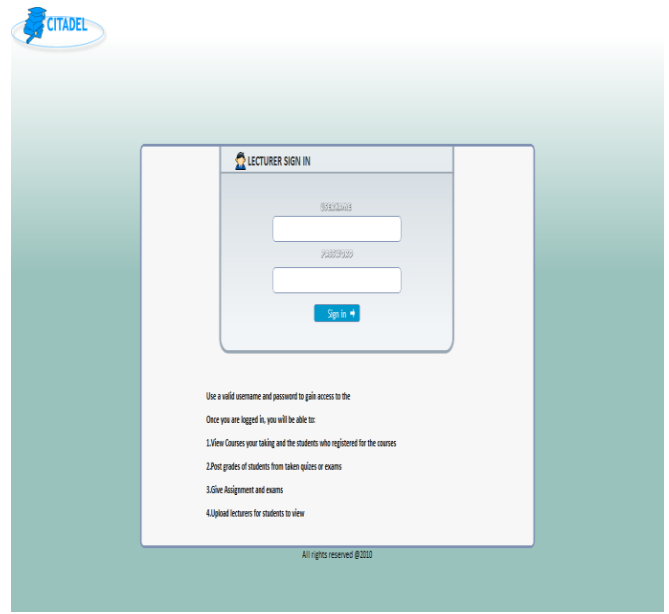

Figure 3: Lecturer's login page

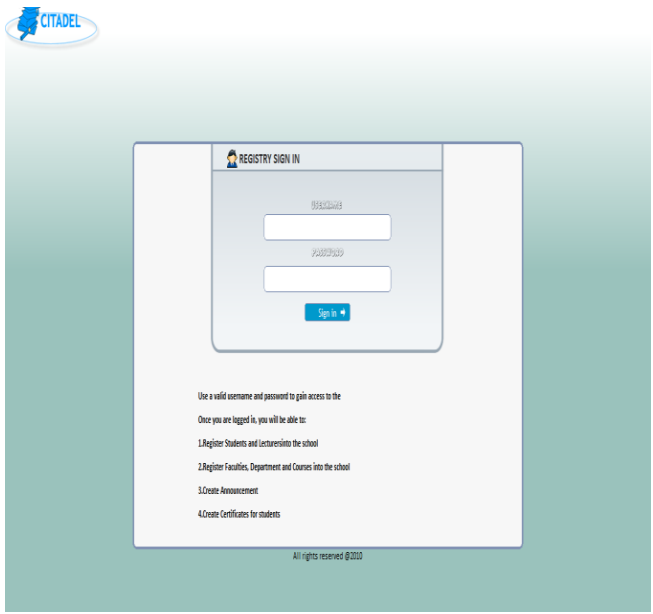

Figure 4: Registrar login page

## V. CONCLUSION

The use of E-learning management systems is still growing so is its acceptance rate in Nigeria as a suitable replacement for traditional learning system. It has advantages that supersede the disadvantages that in no time it will be widely acceptable. The survival and future of tertiary institution greatly depend on E-learning.

The use of the citadel E-learning platform allows institutions foster teaching activity (both by saving administrative and communication efforts) as well as providing the opportunity to improve the learning experience by means of innovative proposals which may enhance the teaching and learning process

## VI. RECOMMENDATIONS FOR FURTHER WORK

This software has attempted to solve the problem of traditional education system and to a large extent it is successful. Regardless of the fact that the system has met the basic objectives of the work, there are ways it could be improved upon and used by other organisations for training or other forms of education sector for usage. It is therefore recommended that further research be carried out on this work to improve it functionality and increase its features. A vital functionality is to ensure that the lecturer, during video conferencing should be able to view physically, the students available for lectures. This will verify the students' attendance. Further research can also be done on how to incorporate the student payment into the study.

## APPENDIX

Below is the list of other windows generated from Citadel E-learning environment:

- **Student home page:** The home page is the first page that the student will encounter after logging in successfully. From the home page the student will be able to navigate to various features of the software.

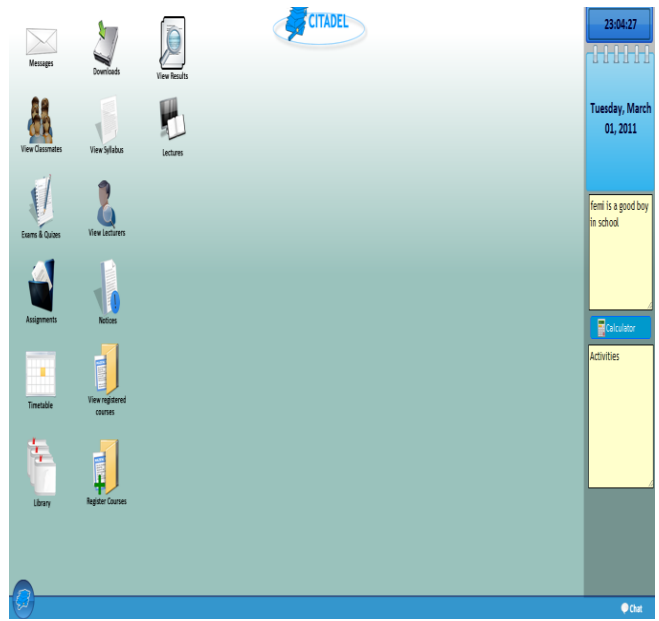

Figure 5: Homepage page

- **Messages**: this feature is for students to receive and send messages to their lecturers and classmates.

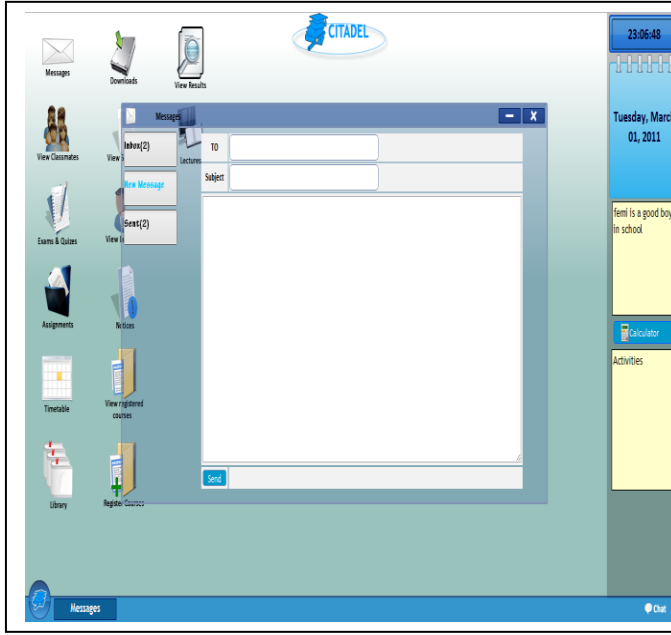

Figure 6: Message window

- **View Classmates**: This feature is for the student to be able to view details such as contact details about other students in their class as depicted below

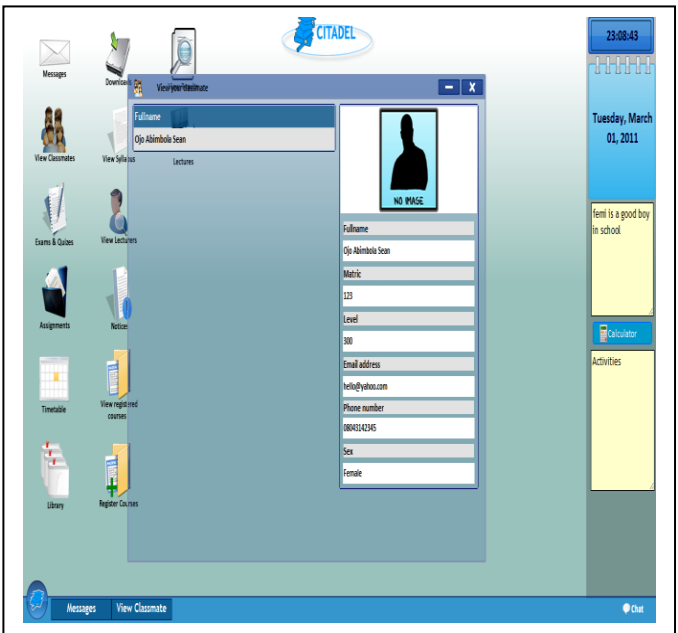

Figure 7: View Classmate window

- **Exams and Quizzes:** This is a section of the program which allows the student to take an exam or quiz as long as they don't miss their deadline.

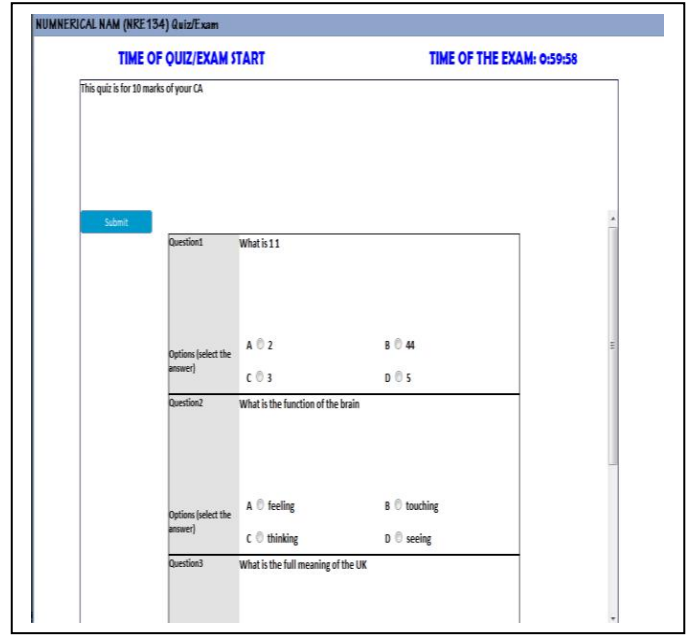

Figure 8: Exam and quizzes window

- **Assignment**: This is a section for viewing or downloading and submitting assignments.

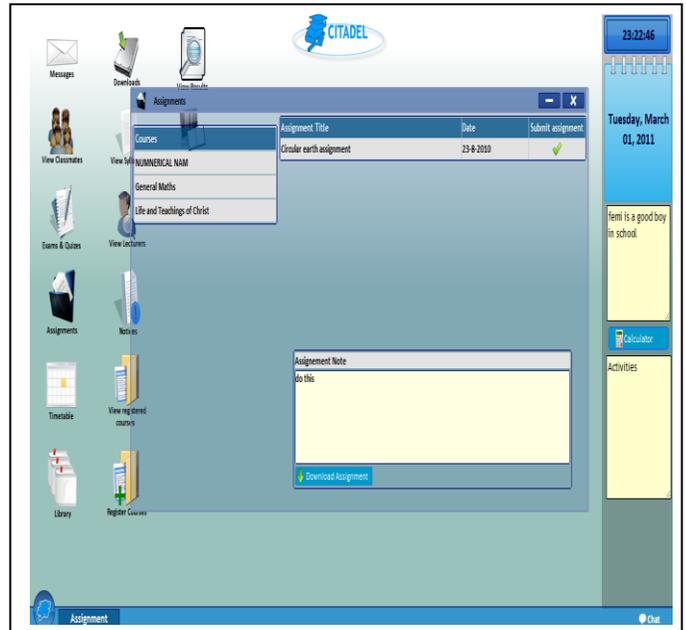

**Figure 9: Assignment window**

- **Timetable:** This is a feature which updates the student about activities such as exams and quizzes of the day.

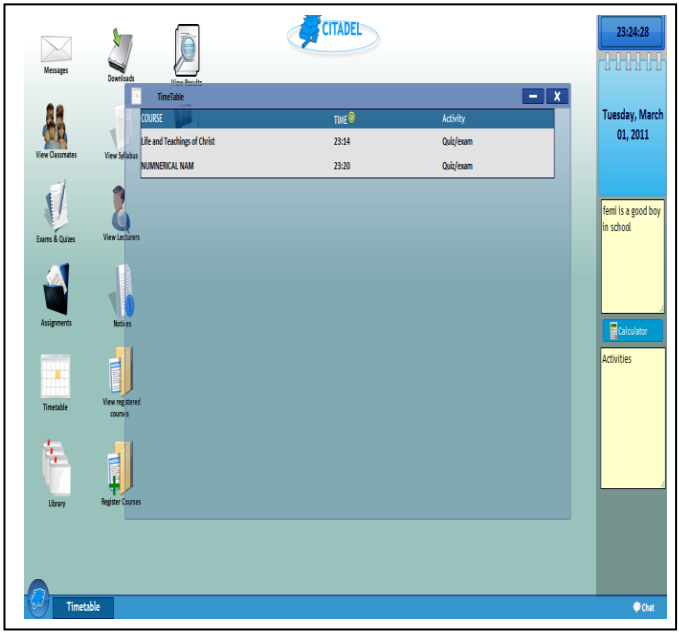

Figure 10: Timetable window

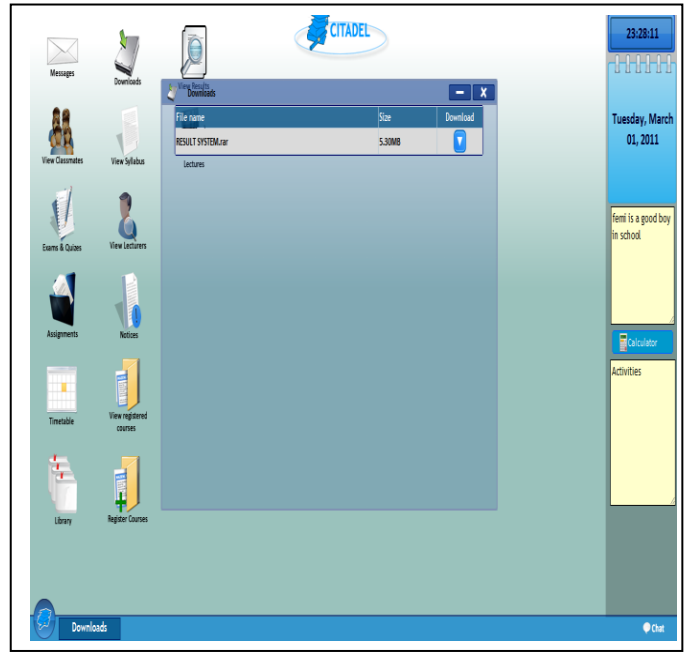

Figure 12: Download Window

- **Library**: As shown in Fig. 11, this is section for viewing available books and details about them in physical library

**View Syllabus:** This section is for viewing syllabus for the courses which the students are taking. This is shown below

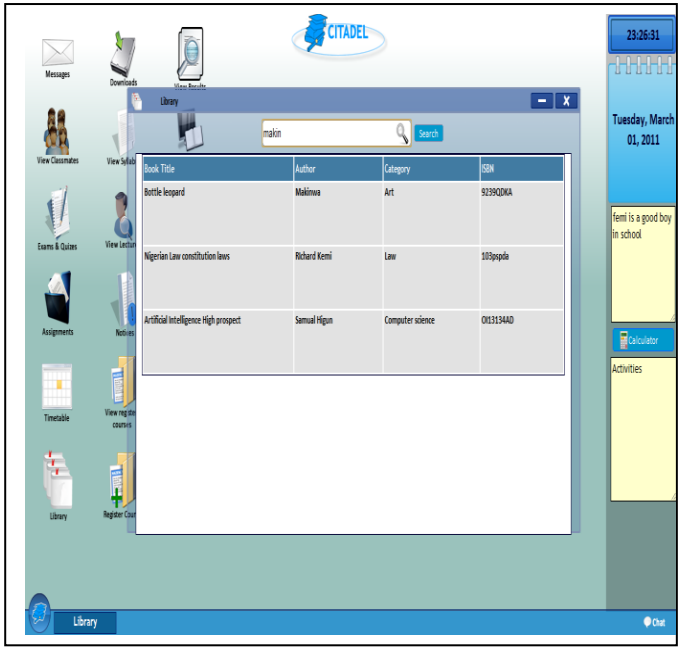

**Figure 11: Library Window**

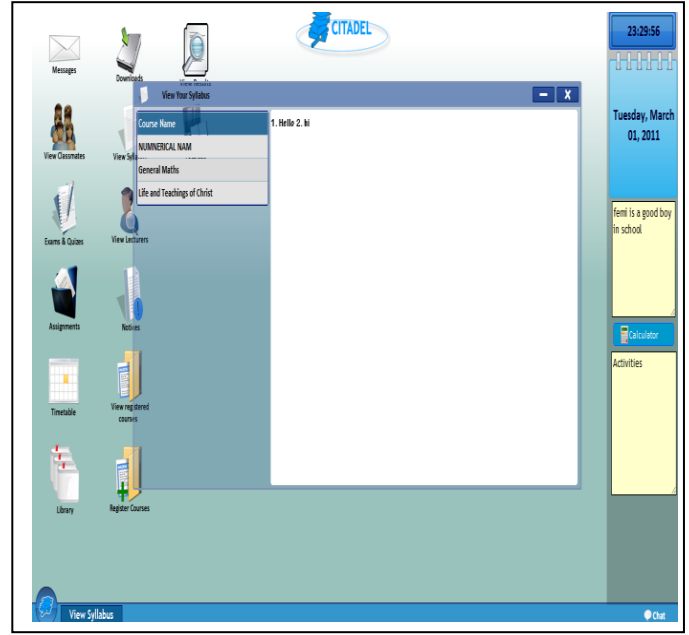

Figure 13: View Syllabus Window

- **Downloads:** This is for downloads which the students might need in the course of using the system to view or access lecture materials.

- **View Lecturers:** This is for viewing details about the lecturers who are taking courses registered by the students.

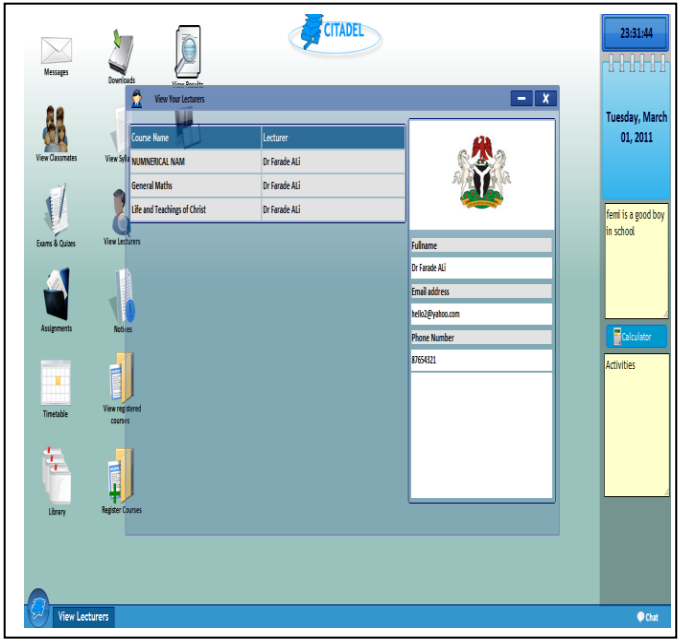

Figure 14: View Lecturers Window

- **Notice:** This is for viewing notices created by the lecturers, registry, library etc.

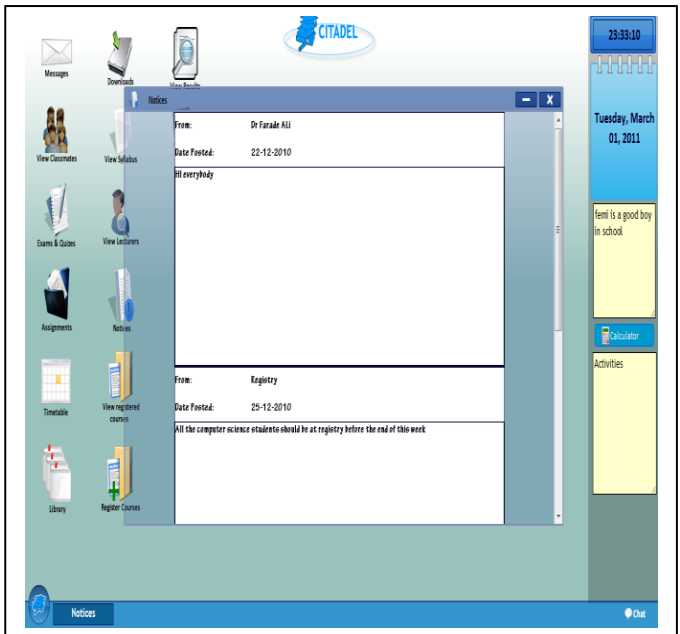

Figure 15: Notice Window

- **Lectures:** This is a section which allows the students to view and download lectures uploaded by the lecturers.

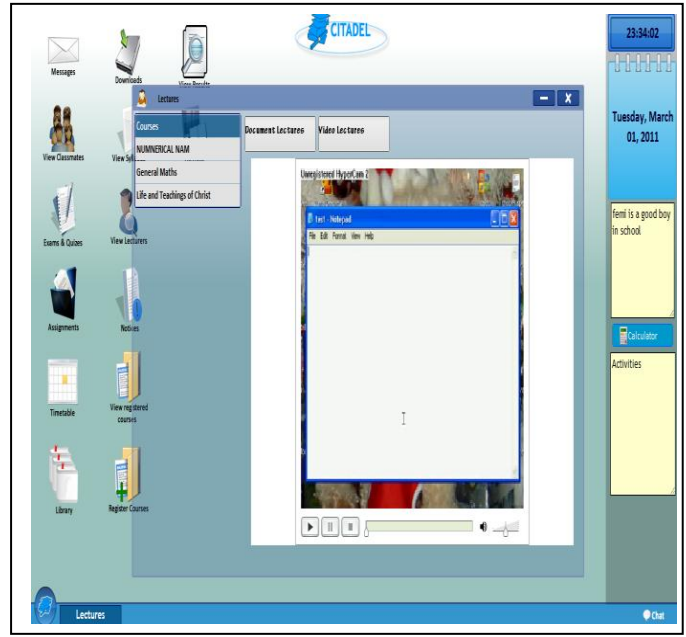

Figure 16: Lectures Window

- **Chat:** This is a section the students use for chatting with the other classmates.

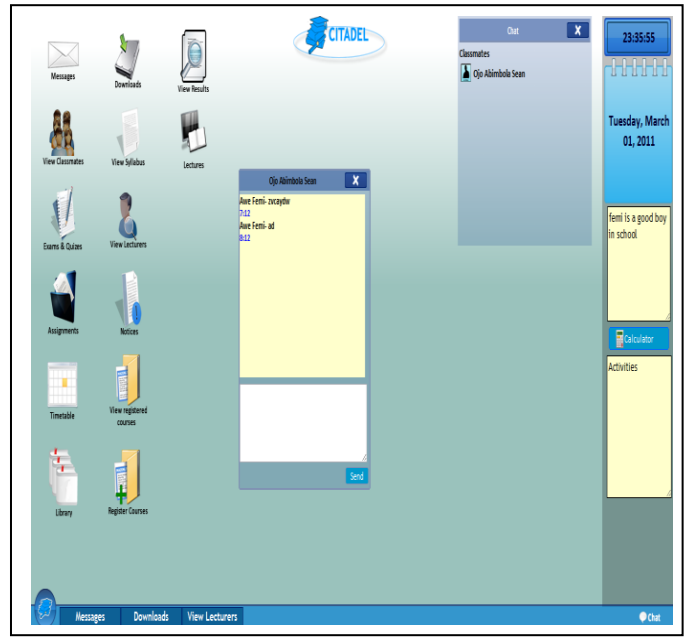

Figure 17: Chat Window

## AUTHORS PROFILE

[1] Dr. Awodele Oludele is a Senior Lecturer at the Computer Science and Mathematics Department, Babcock University Ilishan-Remo, Ogun State, Nigeria. Email: delealways@yahoo.com; Phone: (+234) 8033378761. His research interests are

[2] Ms. S. O. Kuyoro is an assistant lecturer at the Computer Science and Mathematics Department, Babcock University Ilishan-Remo, Ogun State,

Nigeria. Email: afolashadeng1@yahoo.com; Phone: +(234) 8066188970. Her research interests are in the area of data mining and artificial intelligence.

[3] Ms. Adejumobi A. K. is a final year student of the Department of Computer Science and Mathematics, Babcock University, Ilishan-Remo, Ogun State, Nigeria. Email: adetutuadejumobi@gmail.com; Phone: +(234) 7033900451. Her research interests are Database administration; and System Analysis and design.

[4] Mr. Awe O. is a final year student of the Department of Computer Science and Mathematics, Babcock University, Ilishan-Remo, Ogun State, Nigeria. Email: meffess23@yahoo.com; Phone: +(234) 8060987929. His research interest is web application development.

[5] Ms. Makanju O. is a final year student of the Department of Computer Science and Mathematics, Babcock University, Ilishan-Remo, Ogun State, Nigeria. Email: fomak23@yahoo.com; Phone: +(234) 8121152468. Her research interests are web application and software development.